\documentclass[manuscript]{aastex631}

\usepackage{color}
\usepackage{graphicx}
\usepackage{bm}
\usepackage{epsf}
\usepackage{wrapfig}
\usepackage{multirow}
\usepackage{epstopdf}
\usepackage{tabularx}
\usepackage{amssymb,amsmath}

\shorttitle{solar cycle}
\shortauthors{Huang et al.}

\begin{document}

\title{Modeling the Solar Wind During Different Phases of the Last Solar Cycle}

\correspondingauthor{Zhenguang Huang}
\email{zghuang@umich.edu}

\author{Zhenguang Huang}
\affiliation{Climate and Space Sciences and Engineering, University of Michigan, Ann Arbor, MI 48109, USA}

\author{G\'abor T\'oth}
\affiliation{Climate and Space Sciences and Engineering, University of Michigan, Ann Arbor, MI 48109, USA}

\author{Nishtha Sachdeva}
\affiliation{Climate and Space Sciences and Engineering, University of Michigan, Ann Arbor, MI 48109, USA}

\author{Lulu Zhao}
\affiliation{Climate and Space Sciences and Engineering, University of Michigan, Ann Arbor, MI 48109, USA}

\author{Bart van der Holst}
\affiliation{Climate and Space Sciences and Engineering, University of Michigan, Ann Arbor, MI 48109, USA}

\author{Igor Sokolov}
\affiliation{Climate and Space Sciences and Engineering, University of Michigan, Ann Arbor, MI 48109, USA}

\author{Ward B. Manchester}
\affiliation{Climate and Space Sciences and Engineering, University of Michigan, Ann Arbor, MI 48109, USA}

\author{Tamas I. Gombosi}
\affiliation{Climate and Space Sciences and Engineering, University of Michigan, Ann Arbor, MI 48109, USA}



\begin{abstract}

We describe our first attempt to systematically simulate the solar wind during different phases of the last solar cycle with the Alfv\'en Wave Solar atmosphere Model (AWSoM) developed at the University of Michigan. Key to this study is the  determination of the optimal values of one of the most important input parameters of the model, the Poynting flux parameter, which prescribes the energy flux passing through the chromospheric boundary of the model in the form of Alfv\'en wave turbulence. 
It is found that the optimal value of the 
Poynting flux parameter is correlated with the area of the open magnetic field regions with the Spearman's  correlation coefficient of 0.96 and anti-correlated with the average unsigned radial component of the magnetic field with the Spearman's correlation coefficient of -0.91. Moreover, the Poynting flux in the open field regions is approximately constant in the last solar cycle, which needs to be validated with observations and can shed light on how Alfv\'en wave turbulence accelerates the solar wind during different phases of the solar cycle. Our results can also be used to set the Poynting flux parameter for real-time solar wind simulations with AWSoM.

\end{abstract}



\section{Introduction}

The solar wind is a continuous plasma flow expanding from the solar corona and propagating through the heliosphere at supersonic speeds as first proposed by 
\cite{Parker_1958}. Since the time of its prediction, modeling
the solar wind has become an important topic. 
Over the past few decades, various analytical and numerical magnetohydrodynamic (MHD) models of the solar corona have been developed and successfully applied to simulate the background solar wind \citep[e.g.][]{Mikic_1999,Groth_2000,Roussev_2003,Cohen_2007,Feng_2011,Evans_2012}.
Many first-principles models consider Alfv\'en wave turbulence
as the energy source to heat the solar corona and accelerate the solar wind, beginning with early 1D models developed
by \citet{Belcher_1971} and \citet{Alazraki_1971}, to 2D models proposed by \cite{Bravo_1997, Ruderman_1998, Usmanov_2000}, and more recently, 3D models including \cite{Lionello_2009, Downs_2010, vanderHolst_2010}.
Many physical processes associated with the Alfv\'en wave turbulence, such as non-linear
interactions between forward propagating and reflected Alfv\'en waves, are included to
improve the description of coronal heating 
\citep{Velli_1989, Zank_1996, Matthaeus_1999, Suzuki_2006, Verdini_2007, Cranmer_2010, Chandran_2011, Matsumoto_2012}. 
Moreover, heat conduction, radiative losses and energy partitioning among particle species as well as temperature anisotropy 
were introduced in extended MHD (XMHD) models \citep{Leer_1972,Chandran_2011,Vasquez_2003, Li_2004, Sokolov_2013, vanderholst_2014}.
The latest generation of these models is capable of predicting a variety of solar wind observables,
including the solar wind density, velocity, the electron and proton (parallel and perpendicular) temperatures, 
the turbulent wave amplitudes, as well as the wave reflection and dissipation rates, at various locations
in the heliosphere. 

The Alfv\'en Wave Solar atmosphere Model (AWSoM) is one of the commonly used first principles Alfv\'en wave turbulence models developed at the University of Michigan over more than a decade \citep{vanderHolst_2010,Sokolov_2013,Oran_2013,vanderholst_2014,Sokolov_2021}. The model has been extensively validated against observations for solar minimum \citep{Jin_2012, Sachdeva_2019} and maximum conditions  \citep{Sachdeva_2021} including comparisons with recent Parker Solar Probe (PSP) encounters \citep{vanderholst_2019, vanderholst_2022}. 
For these periods of varying solar magnetic activity, the simulated results have been compared to a comprehensive set of observations spanning the low corona to the inner heliosphere. 

The Poynting flux parameter (Poynting flux per B {ratio}) is one of the important inputs for AWSoM, which describes how much Alfv\'en wave energy is entering into the system to heat the corona and power the solar wind into the inner heliosphere. 
\cite{Sokolov_2013} and \cite{vanderholst_2014} estimated this parameter to be approximately $1.1\,\mathrm{MWm}^{-2}\mathrm{T}^{-1}$ based on the chromospheric turbulence observed
by {\it Hinode} \citep{de_Pontieu_2007}.
However the value was modified to $1\,\mathrm{MWm}^{-2}\mathrm{T}^{-1}$
for solar minimum conditions \citep{Sachdeva_2019} and $0.5\,\mathrm{MWm}^{-2}\mathrm{T}^{-1}$ for solar maximum \citep{Sachdeva_2021} conditions
to obtain the best agreement with
both in-situ and remote observations.
It is still unclear how input parameters need to be adjusted to best
simulate the solar wind properties for a specific Carrington rotation. 
This manuscript aims to fill the gap by determining the optimal 
value of one of the important input parameters of the model, the Poynting flux parameter, during different phases
of the solar cycle $24$.
We will also examine the correlation between the optimal value and the underlying physical quantities/processes.

\section{Methodology} 

The detailed description of AWSoM can be found in previous publications \citep{Sokolov_2013,Oran_2013,vanderholst_2014,Sokolov_2021}.
Here we only provide a brief overview.
AWSoM is implemented in the
BATS-R-US (Block Adaptive Tree Solar Wind Roe-type
Upwind Scheme) code \citep{Groth_2000, Powell_1999} within the Space Weather Modeling Framework (SWMF) \citep{Toth_2005, Toth_2012, Gombosi_2021}. 
The model is driven by the observed radial magnetic field component at the inner boundary 
located in the lower transition region with a uniform number density ($2\times10^{17}$\,m$^{-3}$)
and temperature (50,000\,K) distribution.
The underlying assumption is that the  Alfv\'en wave turbulence, its pressure and nonlinear dissipation, is the only momentum
and energy source for heating the coronal plasma and driving the solar wind, 
without considering other potential wave heating mechanisms or contributions from small scale reconnections.
Floating boundary condition is applied at the outer boundary so that the simulated 
solar wind can freely leave the simulation domain.

AWSoM has very few adjustable input parameters. 
The two important input parameters are the Poynting flux parameter, which is
specified as the ratio of the Poynting flux and the magnetic field magnitude at the inner boundary,
and the correlation length of the Alfv\'en wave dissipation (see \cite{vanderholst_2014} and \cite{Jivani_2023}).
The Poynting flux parameter determines the energy input to heat the solar corona and accelerate the solar wind, 
while the correlation length describes how Alfv\'en wave turbulence 
dissipates energy in the solar corona and heliosphere. In this manuscript, 
we focus on the Poynting flux parameter, which is specified at the inner boundary
of AWSoM.

In this study, we simulate one Carrington rotation per year from 2011 to 2019 (the Carrington Rotations and the corresponding magnetogram times are listed in Table\,\ref{tab:cr_list}), 
using the Air Force Data Assimilative Photospheric flux Transport (ADAPT)
Global Oscillation Network Group (GONG) magnetograms \citep{Hickmann_2015}, which are publicly available
on \url{https://gong.nso.edu/adapt/maps/gong}. 
ADAPT maps use a flux transport model to estimate the radial magnetic field in the regions where there are limited or no observations.
There are 12 realizations for each ADAPT map corresponding to 12 different specifications of the supergranulation transport parameters. Currently there's no method to pick the best realization before comparing with observations. Ideally, we should run all 12 realizations for each Carrington rotation and pick the best realization for the corresponding rotation. However, this will increase the computational cost by a factor of 12. With this consideration, we randomly picked the seventh realization for all rotations in this manuscript to reduce the cost of computation. In each Carrington rotation, 
we vary the Poynting flux parameter between 0.3 and 1.2$\,\mathrm{MWm}^{-2}\mathrm{T}^{-1}$ with every 0.05$\,\mathrm{MWm}^{-2}\mathrm{T}^{-1}$ (this range is adjusted to [0.1, 0.95]$\,\mathrm{MWm}^{-2}\mathrm{T}^{-1}$ with every 0.05$\,\mathrm{MWm}^{-2}\mathrm{T}^{-1}$ between [0.2,0.95]$\,\mathrm{MWm}^{-2}\mathrm{T}^{-1}$ and every 0.025$\,\mathrm{MWm}^{-2}\mathrm{T}^{-1}$ below 0.2$\,\mathrm{MWm}^{-2}\mathrm{T}^{-1}$ for CR2137 and CR2154 as the optimal value is either smaller or equal to 0.3$\,\mathrm{MWm}^{-2}\mathrm{T}^{-1}$) to obtain different solar wind solutions and compare the simulated solar wind
with the OMNI hourly solar wind observations. We then calculate the distance between the simulation results and observations
following the methodology introduced by \cite{Sachdeva_2019}, which quantifies the differences between the simulations and in situ observations at 1\,AU to
evaluate the performance of the model. The optimal value of the Poynting flux parameter is chosen when
the simulated solar wind density and velocity are best compared with the observed values, as
these two quantities are most important affecting the CME propagation.
It's important to point out that we limit the model validation to the in-situ data comparison. Other solar corona observations,  for example, the white light images \citep{Badman_2022}, are not included in the current study.

\begin{table}[]
\center
\begin{tabular}{| c | c | c | c |}
\hline
Carrington Rotation & UTC Time of the Magnetogram   & Realization & Optimal Value       \\ \hline
2106                    & 2011-2-2  02:00:00  & 7 & 0.85 \\ \hline
2123                    & 2012-5-16 20:00:00  & 7 & 0.35 \\ \hline
2137                    & 2013-5-28 20:00:00  & 7 & 0.175 \\ \hline
2154                    & 2014-9-2 20:00:00   & 7 & 0.3 \\ \hline
2167                    & 2015-8-23 02:00:00  & 7 & 0.5 \\ \hline
2174                    & 2016-3-3 02:00:00   & 7 & 0.5 \\ \hline
2198                    & 2017-12-17 02:00:00 & 7 & 0.7 \\ \hline
2209                    & 2018-10-13 06:00:00 & 7 & 1.1 \\ \hline
2222                    & 2019-10-2 02:00:00  & 7 & 1.1 \\ \hline
\end{tabular}
\caption{All the ADAPT-GONG magnetograms used in this study and the optimal values of the Poynting flux parameter in the unit of $\,\mathrm{MWm}^{-2}\mathrm{T}^{-1}$.}
\label{tab:cr_list}
\end{table}

\section{Simulation Results} 

\begin{figure}[ht!]
\center

\gridline{\fig{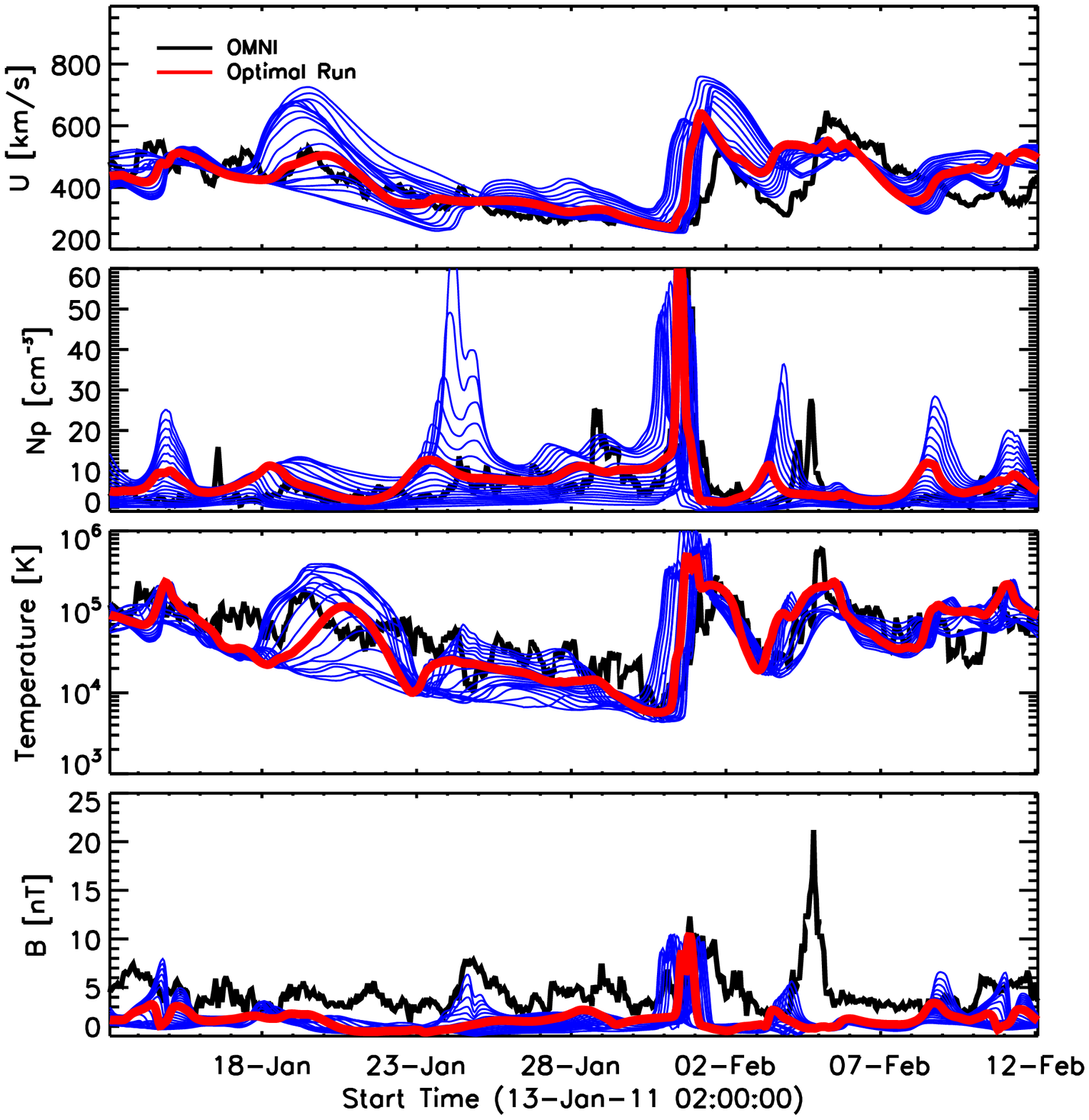}{0.45\textwidth}{(a)}
          \fig{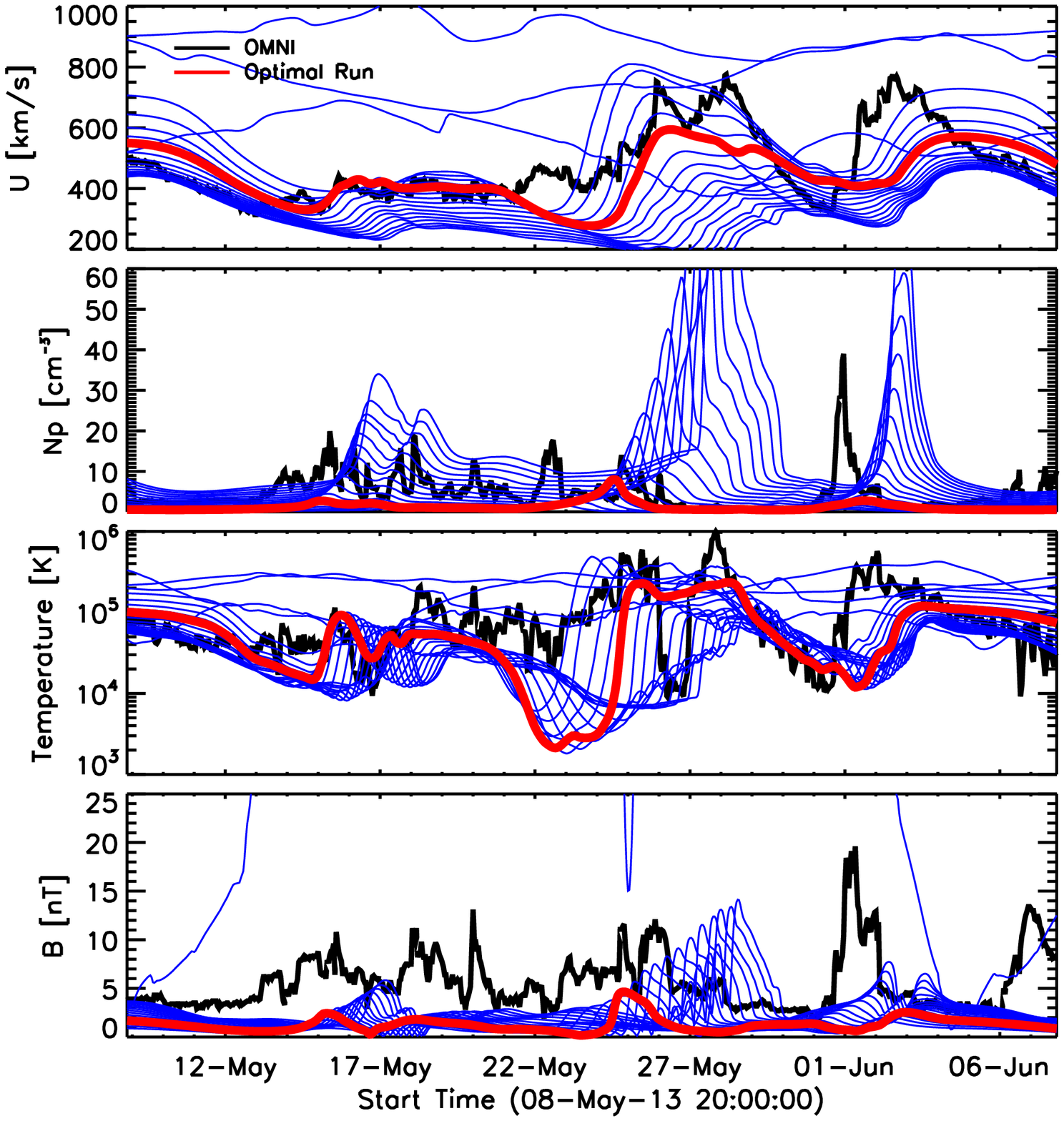}{0.45\textwidth}{(b)}}
          
\caption{AWsoM simulated solar wind bulk velocity, density, temperature and magnetic field (from top to bottom) compared with hourly OMNI observations (black lines) at 1\,AU, for the Carrington rotation 2106 (the left panels, 
a relatively quiet Sun) and 2137 (the right panels, a more active Sun). 
The blue lines correspond to the simulation results with different Poynting flux parameters. 
The red line highlights the results obtained with the optimal value based 
on the best match with the observed solar wind density and velocity. Note the greater variation in plasma quantities for CR2137.}
\label{fig:diff_pf}
\end{figure}

Figure\,\ref{fig:diff_pf} shows the simulated solar wind for two Carrington rotations, one in 2011 (CR2106) near solar minimum and the other in 2013 (CR2137) near solar maximum. 
Each blue line represents one AWSoM simulation result with a given Poynting flux parameter (while all
other parameters are kept the same). The red lines highlight the best run in the corresponding rotation, 
based on the the best comparison with the observed solar wind density and velocity. 
The plots illustrate that different values of the Poynting flux parameter can drastically change the simulated
solar wind at 1\,AU for an active Sun (in 2013).
Many simulations give unreasonable results, e.g. very large magnetic field ($>$25 nT) or solar wind velocity far from observations. 
Close to the solar minimum in 2011, the Poynting flux
parameter has much less impact, but it is still causes significant variations. In both cases, 
it is critical to use the correct parameter, otherwise the simulation results will be incorrect.

\begin{figure}[ht!]
\center

\gridline{\fig{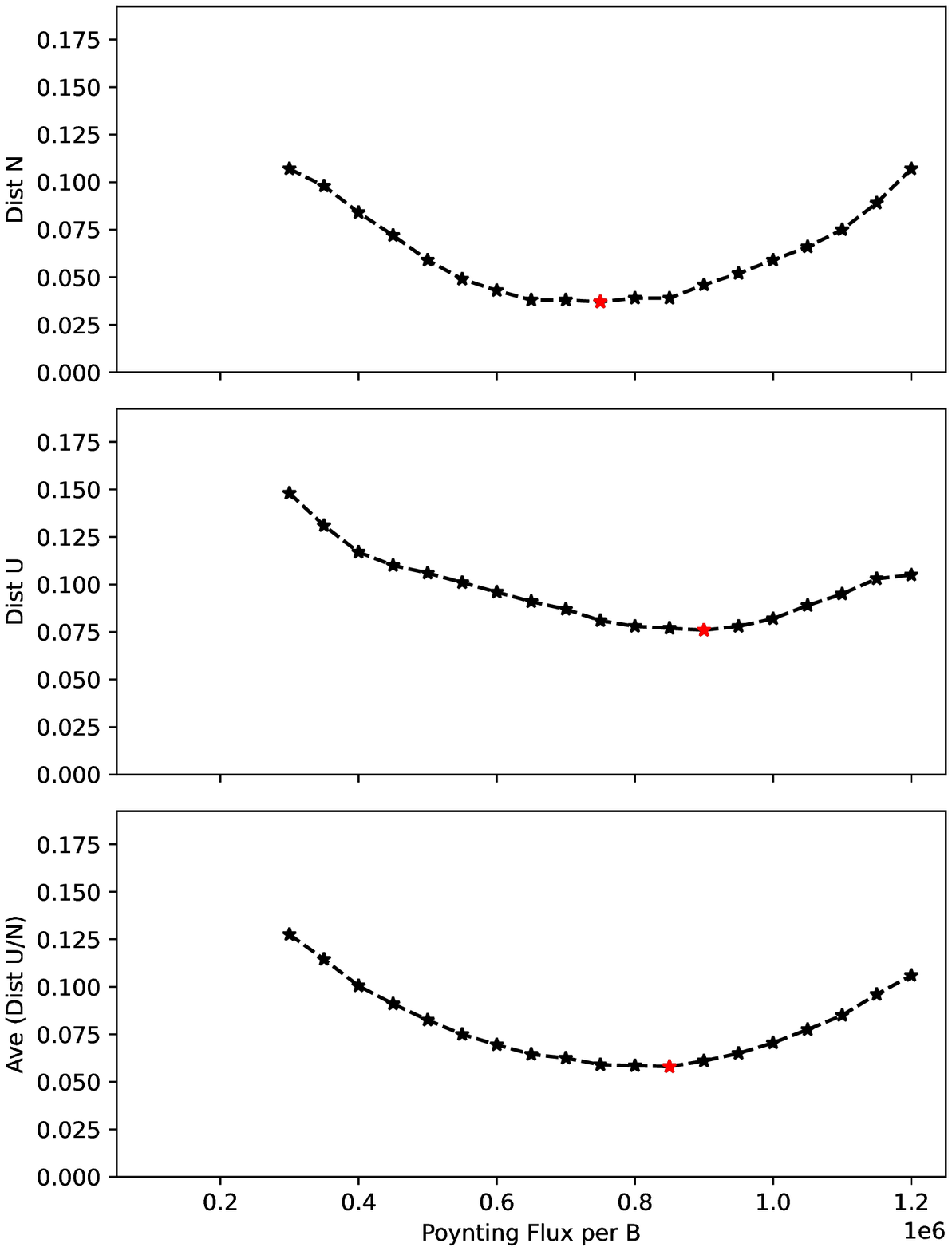}{0.4\textwidth}{(a)}
          \fig{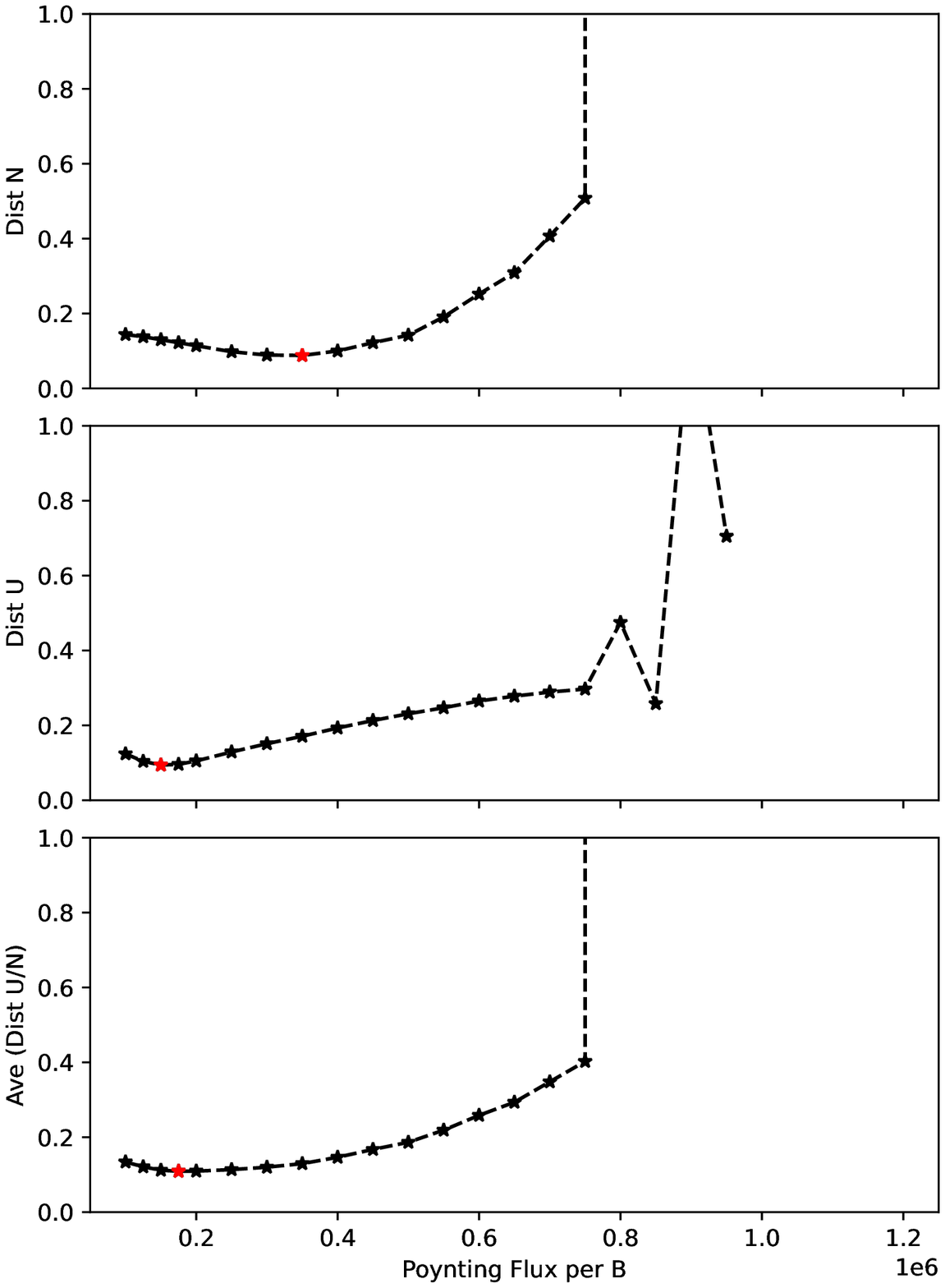}{0.4\textwidth}{(b)}}
          
\caption{The distances between the simulated and observed solar wind for CR2106 (the left panel) and CR2137 (the right panel). 
The x-axis shows the value of the Poynting flux parameter
while the y-axis plots the distances between the simulations and OMNI observations. 
The top row shows the distances between the 
simulated and observed densities, while
the second row shows the distances between the 
velocities. The third row displays the average
of velocity and density distances.
The optimal Poynting flux parameter (colored with red) 
for each panel is found at the minimum of the curves.}
\label{fig:dist}
\end{figure}

The differences between the simulated and observed solar wind time-series data are quantified by the distances proposed by
\cite{Sachdeva_2019} and were used to determine the best comparison with observation. We calculate the distances between the simulated and observed densities as well as velocities, 
as these two quantities significantly impact the CME propagation.
We also calculate the average of the density and velocity distances to describe the overall performance.  Figure\,\ref{fig:dist}
shows that the optimal value of the Poynting flux parameter depends on the choice of the error criteria (for example, based on the density or velocity distances). 
In this study, we select the optimal value of the Poynting flux parameter when the
average distance of the density and velocity reaches the minimum value. Figure\,\ref{fig:dist}
suggests a monotonic increase of the distance when the Poynting flux parameter is smaller
or larger than the optimal value, which means that the optimum is reliably defined.  
For CR2137 (near solar max), the simulation results become unrealistic when the Poynting flux parameter is larger than $0.75\,\mathrm{MWm}^{-2}\mathrm{T}^{-1}$ as the distances are very large, which confirms that it is critical to choose a correct the Poynting flux parameter in order to obtain reasonable results.

\begin{figure}[ht!]
\center

\gridline{\fig{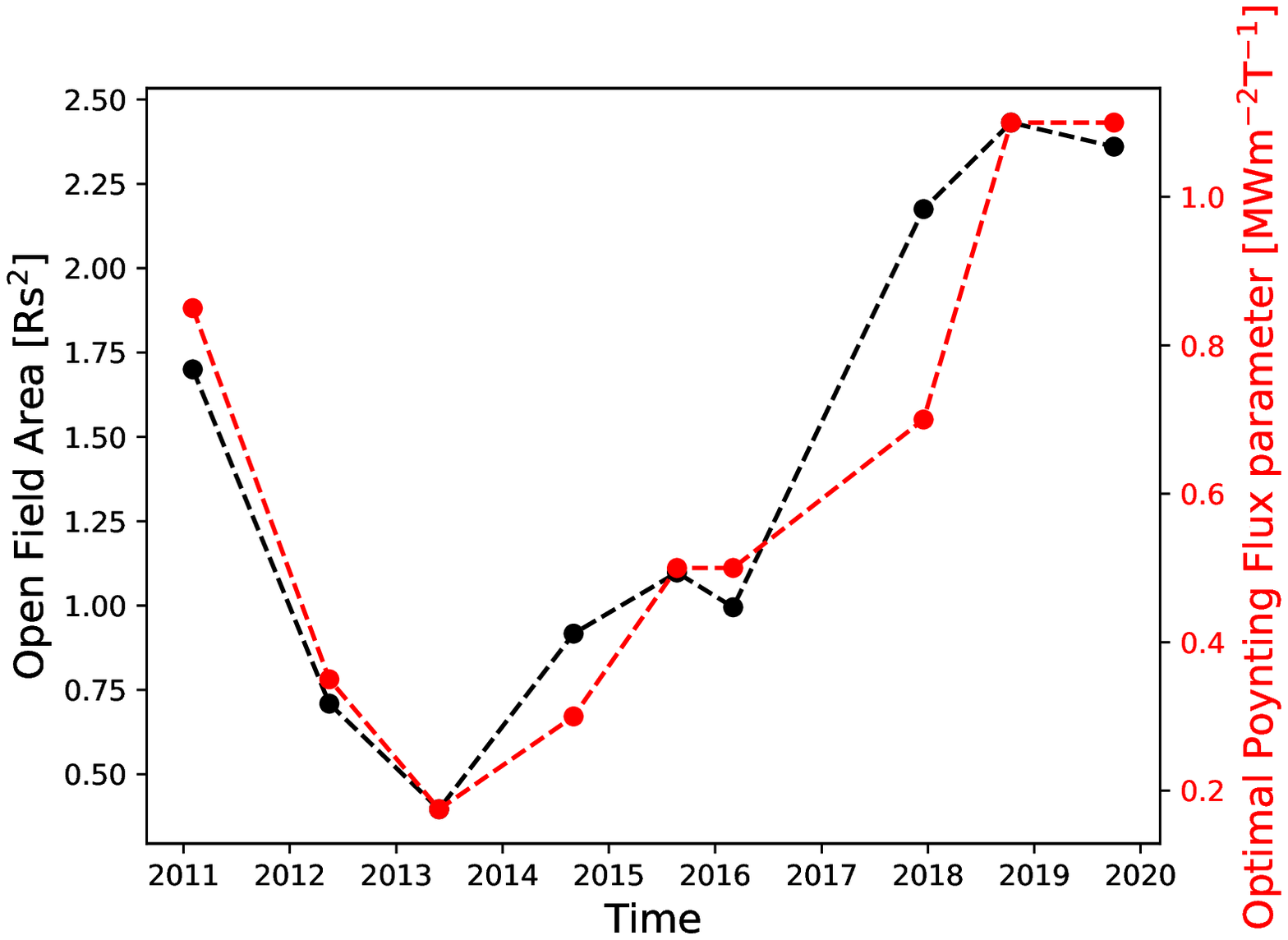}{0.45\textwidth}{(a)}
          \fig{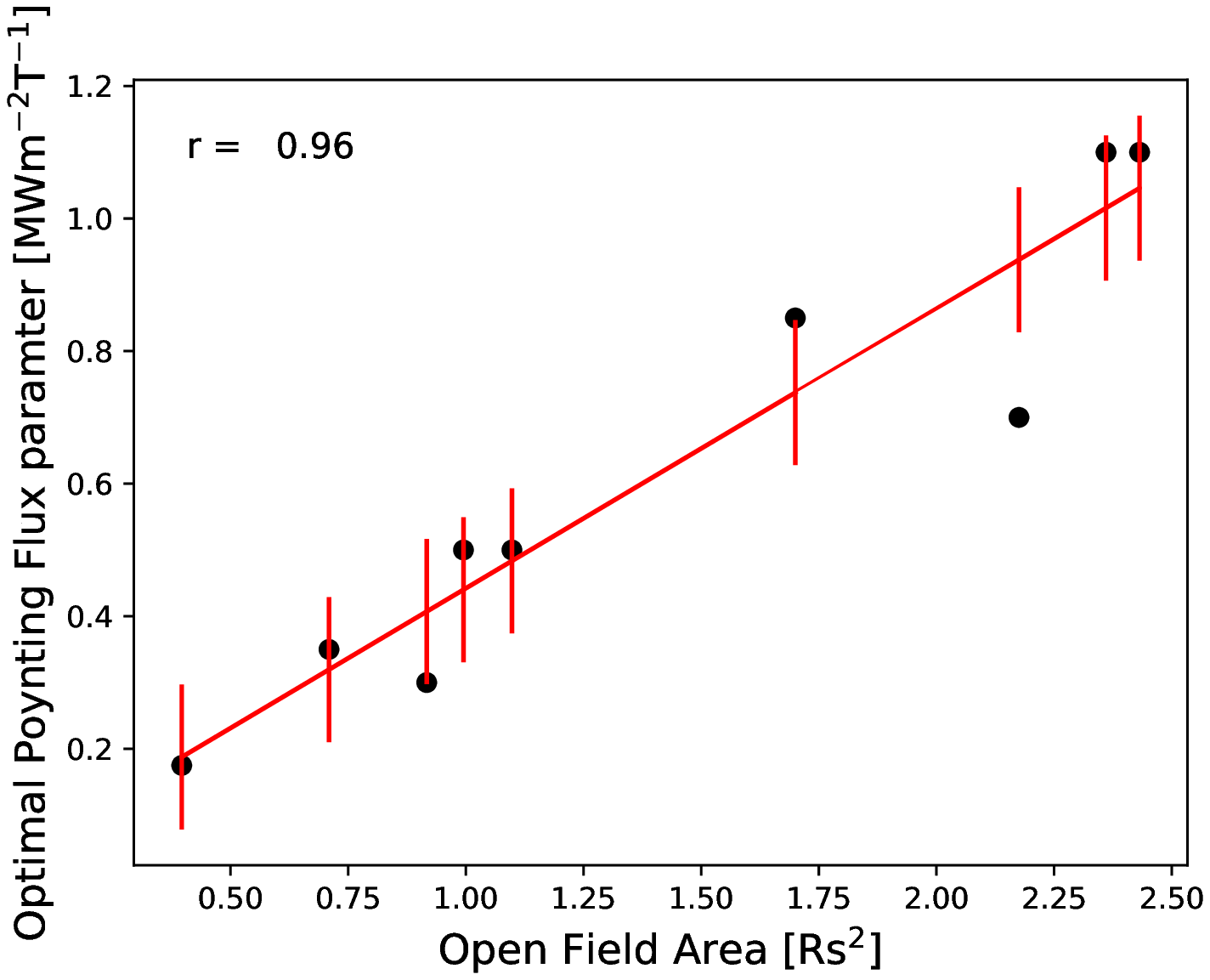}{0.45\textwidth}{(b)}}
\gridline{\fig{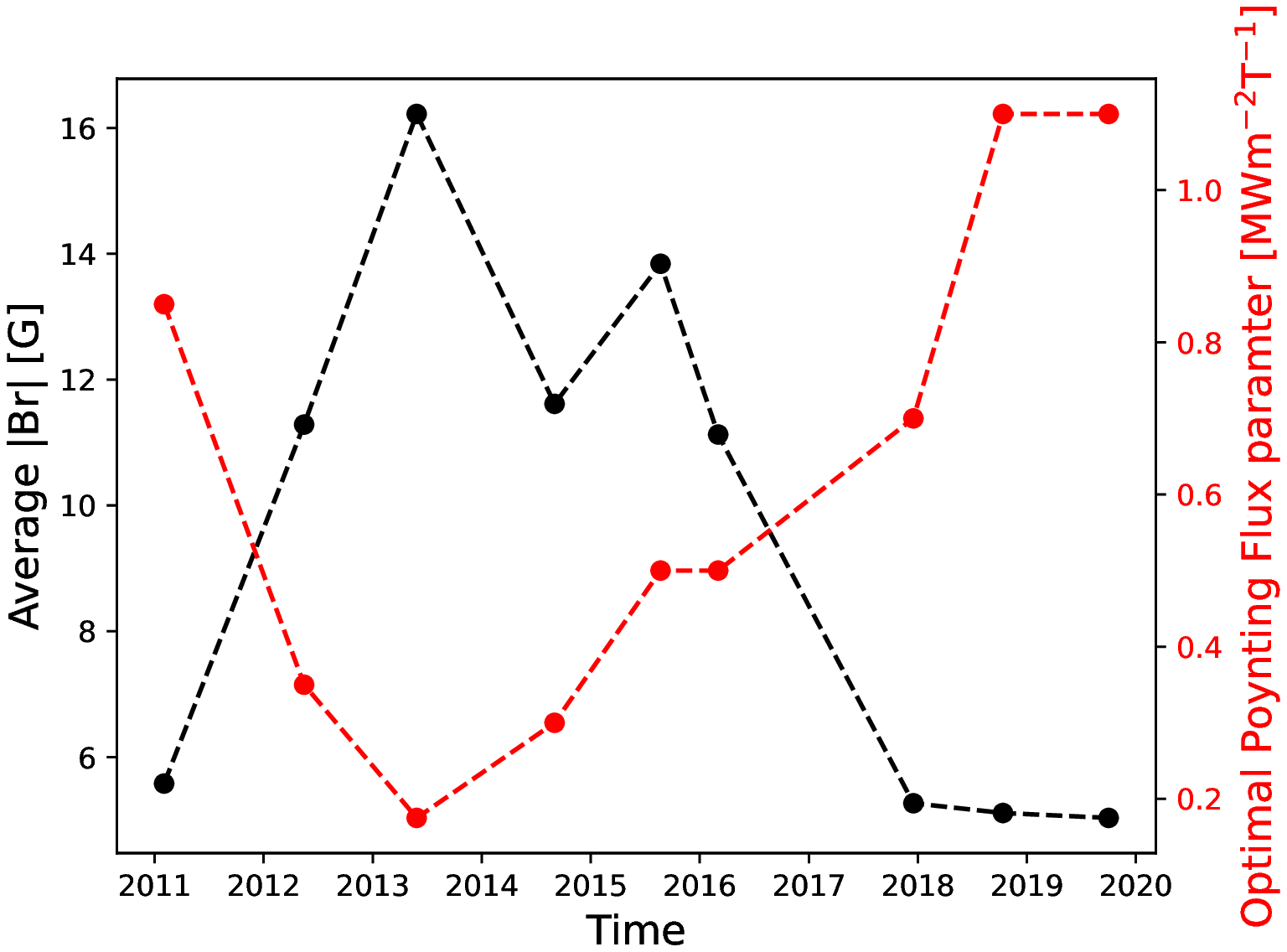}{0.45\textwidth}{(c)}
          \fig{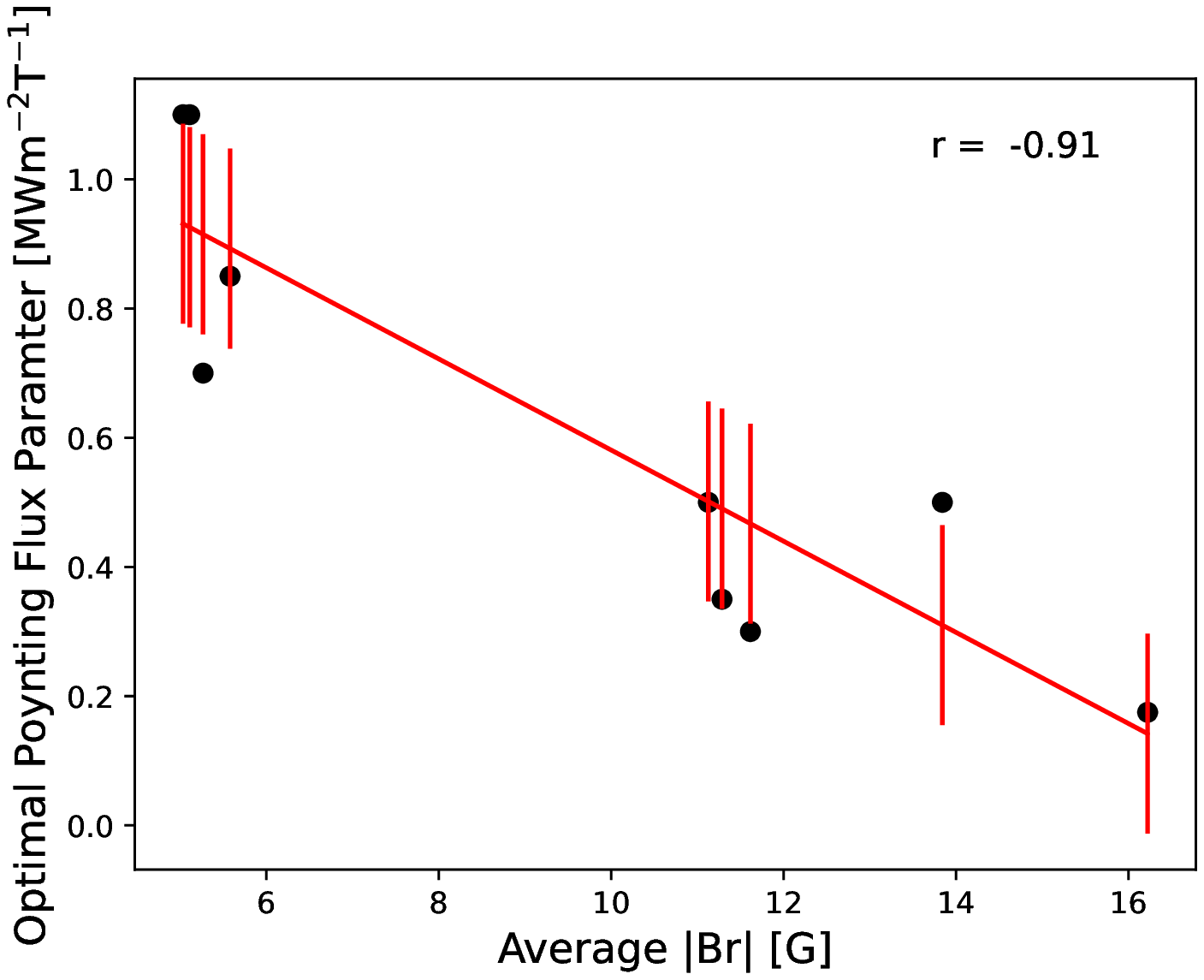}{0.45\textwidth}{(d)}}

\caption{Panel (a) shows the open magnetic field area (black, left axis) and the optimal Poynting flux parameter (red, right axis) as a function of the times of the Carrington rotations. 
Panel (b) shows the Spearman's correlation coefficient $r$ and the linear regression between the area of the open field regions and the optimal Poynting flux parameter with error bars indicating the standard error.
Panel (c) shows the average unsigned $|B_r|$ in the open field regions (black, left axis) and the optimal Poynting flux parameter (red, right axis) as a function of the times of the Carrington rotations.
Panel (d) shows the Spearman's correlation coefficient $r$ and the linear regression between the average $|B_r|$ and the optimal Poynting flux parameter with error bars indicating the standard error.}
\label{fig:stat}
\end{figure}

Table\,\ref{tab:cr_list} lists the optimal value of the Poynting flux parameter for each of the Carrington rotation in this study. A natural question is whether we can predict the optimal Poynting flux parameter without performing dozens of simulations. The
magnetic field structure of the solar corona, for example, may contain some clues. To answer this question, we explore
the relationship between the optimal Poynting flux parameter and  quantities
associated with the magnetic field configurations, including the open magnetic flux (the surface integral of the magnitude of the radial component of the magnetic field $|B_r|$ in the open field regions at the inner boundary), 
the average $|B_r|$ on the whole solar surface or in the open field regions,
and the area of the open magnetic field regions. We find that the optimal Poynting flux parameter
is highly correlated with the area of open field regions (see Panel (b) in Figure\,\ref{fig:stat}) with 0.96 Spearman's correlation coefficient and anti-correlated with the average $|B_r|$ in the open field regions (see Panel (d) in Figure\,\ref{fig:stat}) with -0.91 Spearman's correlation coefficient.
Panels (a) and (c) in Figures\,\ref{fig:stat} show that the optimal Poynting flux parameter and 
the area of the open field regions are anti-correlated with the Sun's activity:
the values of the Poynting flux parameter are small during the solar maximum (around 2013-2014) and then increase towards 
the solar minimum; while  the average unsiged $B_r$ in the open field regions is orrelated with the Sun's activity. We performed a linear regression between the optimal Poynting flux parameter $P$ [MWm$^{-2}$T$^{-1}$] and the open field area $A$ [R$_s^2$] as well as the average unsigned  $B_r$ in the open field regions $B$ [G], and obtained the following following formulas:
\begin{eqnarray}
P &=& 0.42  \cdot A + 0.02 \pm 0.11 \\
P &=& -0.07 \cdot B + 1.29 \pm 0.16
\end{eqnarray}
The $\pm$ terms indicate the standard error of the linear regression.

\section{Summary and Discussions}

Solar wind models based on first principles often assume that Alfv\'en
wave turbulence is the primary energy source to heat the solar corona and accelerate the solar wind. 
All first principles models need input parameters, which are based on either theoretical expectation or observations.
It is important to understand what the physical implications of the input parameters are 
and how these input parameters would need to be adjusted under different solar conditions,
to better understand how the solar corona is heated and the solar wind is accelerated
during a full solar cycle, especially if the theory could self-consistently explain the acceleration mechanism. In this study,
we use AWSoM, which is based on the Alfv\'en wave turbulence theory, to simulate
the solar wind background during different phases of the last solar cycle, and explore
how the input parameters need to be adjusted for different solar conditions.

We found that the optimal Poynting flux parameter, which is determined by minimizing the difference between the
simulated and observed (by OMNI) solar wind densities and velocities, is
highly correlated with the magnetic field structure of the solar corona. To be specific, 
the open magnetic flux and the area of the open field regions are well correlated with the optimal value of the Poynting flux parameter. The solar cycle dependence of the 
area of the open field regions found in our study are consistent with \cite{Nikolic_2019} and \cite{Lowder_2017}. 
On the other hand, the variation of the optimal Poynting flux parameter, which is defined as the ratio of the Poynting flux and the magnetic field magnitude at the inner boundary, is a new result. 
Prior work assumed
a constant value around $1.1\,\mathrm{MWm}^{-2}\mathrm{T}^{-1}$ \citep{Sokolov_2013,vanderholst_2014},  
based on the chromospheric turbulence observed by {\it Hinode} \citep{de_Pontieu_2007}.
However, \cite{de_Pontieu_2007} is a single observation and we are not aware of any study of the chromospheric turbulence during different
phases of the solar cycle. Our study predicts that the chromospheric turbulence may vary during the solar cycle and it's anti-correlated with the average unsigned $B_r$ in the open field regions. 

\begin{figure}[ht!]
\center
\includegraphics[width=0.9\linewidth]{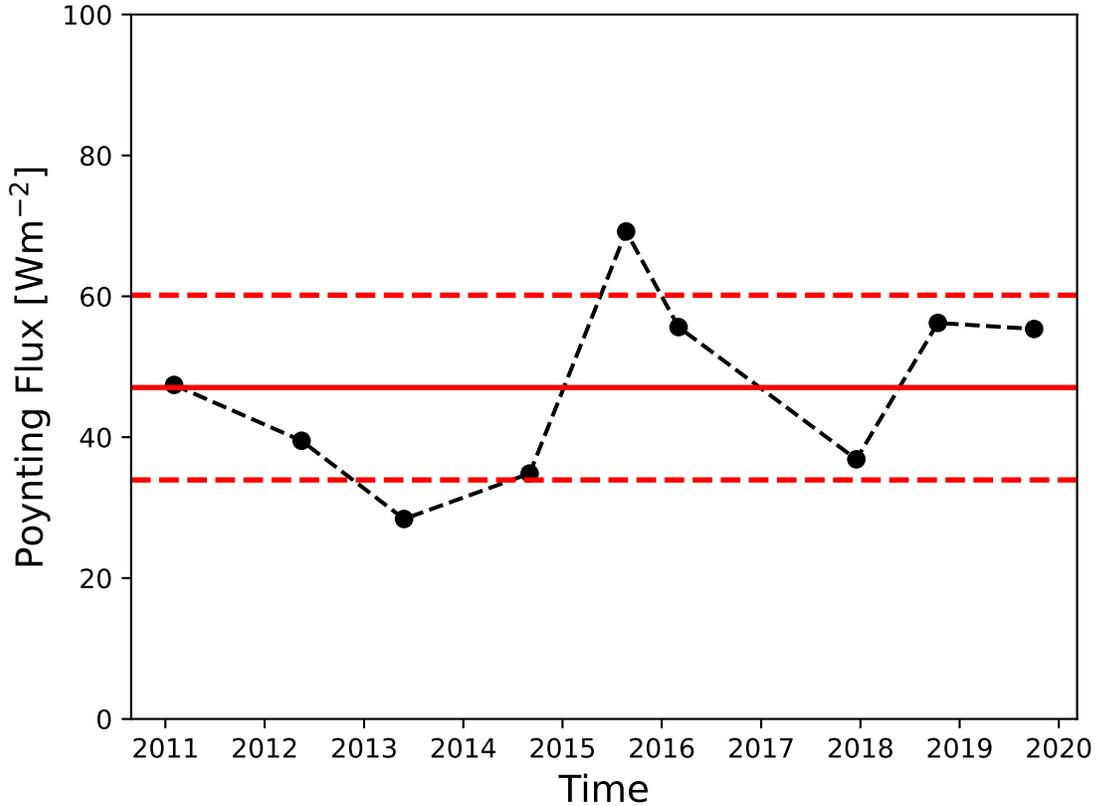}
\caption{The average Poynting flux in the open field regions during the last solar cycle. The red horizontal line shows the average value (47.42\,Wm$^{-2}$) of the Poynting flux and the red dashed horizontal lines indicate the standard deviation $\pm 13.12\,$Wm$^{-2}$.}
\label{fig:PoyntingFlux}
\end{figure}

Figure\,\ref{fig:PoyntingFlux} shows the average Poynting flux in the open field regions, which is the product of the the average unsigned $B_r$ and the optimal Poynting flux parameter (defined as the ratio of the Poynting flux and the magnetic field magnitude) that AWSoM needs to provide the best comparison with OMNI observations. It shows that the variation of the average Poynting flux in the open field regions is significantly smaller than the variations of the Poynting flux parameter and the open field areas. It will be interesting to see if observations confirm (or contradict) our predictions.  A theoretical explanation of why the average energy deposit rate in the open field regions does not change significantly in a solar cycle, as suggested in Figure\,\ref{fig:PoyntingFlux}, could significantly improve our understanding how the Alfv\'en wave turbulence heats the solar corona and accelerates the solar wind in different phases of the solar cycle.

This study is also important for the space weather prediction community.  
First-principles solar wind models have not been used in a real time solar wind prediction primary due to
two reasons: 1. high computational cost; 2. the uncertainty of the input parameters. Nowadays, the
rapid development of supercomputers (e.g., the Frontera system supported by NSF and the 
NASA supercomputing system Pleiades) makes it possible to use a first principles solar wind model to
perform real time solar wind predictions, if the input parameters of the model could be specified correctly. 
The results presented here prescribe one of the important input parameters of AWSoM, 
the Poynting flux parameter, based on the strong correlation with the open magnetic flux and the area of the open field regions. Both of these quantities can be easily obtained with the required accuracy from the potential field source surface model, for example. As the solar cycle 25 approaches, it will be very helpful to investigate if such behavior remains valid. It is also helpful to check if this empirical relation is valid for different solar cycles. Besides, it's unclear if a similar empirical relation is valid for different types of magnetograms. This is a first attempt in this direction and much more work with involvement of different types of magnetogram and additional different solar cycles is needed in the future.

There are a few limitations of the current study. First of all, the study is limited to ADAPT-GONG magnetograms. Previous studies \citep{Jin_2022, Linker_2017, Riley_2021, Perri_2022, Sachdeva_2023}  showed that different magnetograms generally produce different simulated solar wind. Whether the empirical relation could be directly applied to other magnetograms is beyond the scope of this study. We plan to expand the study for different types of magnetograms in the future. 

The study may have some uncertainties during solar maximum. The topology changes dramatically at solar maximum, the simulations sometimes cannot produce good comparisons between the simulations and observations (Panel (b) in Figure\,\ref{fig:diff_pf}), which may be caused by the limitation of the observations: the photospheric magnetic field is most reliable near the center of the solar disk and it may change significantly when it moves to the limb or back of the Sun in a few days. Large uncertainties are then introduced when constructing a synoptic or synchronic magnetogram for a full Carrington rotation. Consequently, the simulated solar wind will have larger uncertainties compared to solar minimum. We plan to study more rotations near solar maximum in the near future to quantify this effect.

\begin{acknowledgments}

This work was primarily supported by the NSF PRE-EVENTS grant No. 1663800, 
the NSF SWQU grant No. PHY-2027555, the NASA grant No. 80NSSC23K0450, and the NASA Heliophysics DRIVE Science
Center (SOLSTICE) at the University of Michigan under grant
NASA 80NSSC20K0600. W. Manchester was also supported by the NASA
grants NNX16AL12G and 80NSSC17K0686. L. Zhao was also supported by the NASA grants 80NSSC21K0417 and 80NSSC22K0269.

We acknowledge the high-performance computing support from Cheyenne
(doi:10.5065/D6RX99HX) provided by NCAR's Computational and Information Systems Laboratory,
sponsored by the NSF, and the computation time on Frontera (doi:10.1145/3311790.3396656)
sponsored by NSF and the NASA supercomputing system Pleiades. 

This work utilizes data produced collaboratively between AFRL/ADAPT and NSO/NISP.

\end{acknowledgments}

\bibliography{reference}

\end{document}